\documentclass{jpp}
\usepackage{graphicx}
\usepackage{epstopdf, epsfig, color}
\newcommand{\il}{~}

\shorttitle{Morphology of the two-dimensional MRI in
Axial Symmetry}
\shortauthor{G. Montani and D. Pugliese}

\title{ Morphology of the two-dimensional MRI in Axial Symmetry}

\author{G. Montani\aff{1}\aff{2}
  \corresp{\email{giovanni.montani@frascati.enea.it }},
  \and D. Pugliese\aff{3}}

\affiliation{\aff{1}ENEA, Unit\`a Tecnica Fusione, ENEA C. R. Frascati, via E. Fermi 45, 00044 Frascati (Roma), Italy
\aff{2}Physics Department, ``Sapienza'' University of Rome, P.le Aldo Moro 5, 00185 (Roma), Italy
\aff{3}Institute of Physics, Faculty of Philosophy \& Science,
  Silesian University in Opava,
 Bezru\v{c}ovo n\'{a}m\v{e}st\'{i} 13, CZ-74601 Opava, Czech Republic}

\begin{document}

\maketitle

\begin{abstract}
In this paper, we analyze the
linear stability of a stellar
accretion disk, having a stratified
morphology. The study is performed in
the framework of ideal magneto-hydrodynamics and therefore it results in a
characterization of the linear unstable
magneto-rotational modes.
The peculiarity of the present scenario
consists of adopting the magnetic flux
function as the basic dynamical variable.
Such a representation of the dynamics
allows to make account of the
co-rotation theorem as a fundamental
feature of the ideal plasma equilibrium,
evaluating its impact on the perturbation
evolution too.
According to the Alfvenic nature of
the Magneto-rotational instability,
we consider an incompressible plasma
profile and perturbations propagating
along the background magnetic field.
Furthermore, we develop a local
perturbation analysis, around fiducial
coordinates of the background configuration and dealing with very small scale
of the linear dynamics in comparison
to the background inhomogeneity size.
The main issue of the present study
is that the condition for the emergence
of unstable modes is the same in the
stratified plasma disk, as in the
case of a thin configuration.
Such a feature is the result of the
cancelation of the vertical derivative
of the disk angular frequency from the
dispersion relation, which implies
that only the radial profile of the
differential  rotation is
responsible for the trigger of growing modes.

\end{abstract}

\section{Introduction}

The request for the existence of a
linear unstable mode spectrum in a
two-dimensional axially symmetric
configuration clearly emerges from the
theory of accretion on a compact astrophysical object. In fact, the Shakura idea
\cite{S73,SS73} that the angular momentum transport
across a stellar accretion disk is
realized via an effective viscosity,
naturally leads to search for a instability
mechanism, able to trigger turbulence in
the plasma profile. In the case of a
differentially rotating disk, embedded in the
central object gravitational field only,
the axial symmetry of the configuration
prevents that any linear unstable mode
arises   \cite{B01}.

However, as a weak magnetic field is
involved in the background disk morphology, the Velikhov-Chandrasekhar instability
\cite{V59,C60},
also known as Magneto-rotational instability
(MRI), is triggered.
The basic mechanism underlying the emergence of such kind of instability is the
direct coupling between the differential
rotation of the disk and the magnetic
field tension. By other words,
the plasma disk inhomogeneity transform
Alfvenic waves into growing modes.

The detailed nature of the MRI and
its role in triggering turbulence across
an accretion disk has been extensively stated
in \cite{BH98}. The condition for
getting the emergence of growing modes
is that the Alfven frequency
(the wavenumber amplitude times the
Alfven speed) is small compared with
a frequency-like term containing
the radial gradient of the disk angular
velocity.

In \cite{Balbus95}, the study of MRI is extended to
a stratified disk (characterized by an inhomogeneous
thick profile), outlining the role that the vertical
derivative of the angular velocity plays in the morphology
of the unstable modes. The study suggests that the
driving force of the instability is to be identified in
the spatial gradient of the angular velocity, differently
from the non-magnetized case, where the dominant
contribution  comes from the specific angular momentum
gradients. We stress how the analysis of the stratified disk,
performed in \cite{Balbus95} relies on a vector formulation of
the Magneto-hydrodynamics (MHD) ideal equations,
i.e. no use is made of the magnetic flux surface function
as a dynamical variable. Furthermore, this study
does not rely on the validity of the co-rotation theorem,
disregarded in the construction of the dispersion relation.

In the present paper, we perform a study of the stratified
disk, similar to the one in \cite{Balbus95}, but based
on the magnetic surface function dynamics and
directly accounting for the co-rotation theorem (the background disk angular frequency
must depend on the flux surface \cite{F37}),  holding
for the background magnetic
configuration. Without a significant loss of generality,
we simplify the analysis of the dispersion relation for
the linear mode spectrum, by considering a vanishing
azimuthal component of the background magnetic field
(as typically true for the magnetic field of a compact
astrophysical object) and we also take perturbations which
propagate along the background magnetic field only
(this feature is intrinsic for Alfvenic disturbances
of the background, as MRI results to be).
More specifically, we deal with a local perturbation scheme,
based on the construction of linear dynamics nearby
fiducial background coordinates, as allowed by the
assumption that the perturbation have a sufficiently short
wavelength to explore a limited portion of the steady profile.
Here, we consider an incompressible plasma at any order
of approximation, coherently with the so-called
Boussinesq approximation \cite{BH98}
(which states incompressible perturbations as a
consequence of the mass conservation equation, when
the large wavenumber hypothesis is implemented).

The main issue of our analysis is showing how the vertical
derivative of the disk angular velocity cancels out
from the dispersion relation, as far as the features of the
co-rotation theorem are retained in the perturbation scheme.
Indeed, our dispersion relation implies the same
morphology of the MRI, (i.e. the same condition
on the background parameters in order to trigger the instability), exactly like in the thin disk scenario.

This fact has a relevant physical implication for the
accretion mechanism onto a compact object, since it states
that only the radial differential rotation of the disk
accounts for the instability property of the plasma
and hence, only the radial steady disk profile
really matters when the transport processes are
analyzed.

The paper is organized as follows. In Sec.\il(\ref{Sec:II})
we provide the general ideal MHD scheme to describe the plasma
disk evolution in the formalism of the magnetic surface functions.
All the basic equations are provided and the main
implications of their structure are traced.
In Sec.\il(\ref{Sec:mor}), we briefly describe the background
plasma configuration, setting the basic force balance
relations. In Sec.\il(\ref{IV}), we develop the linear
perturbation equations, as written in the Fourier
(plane wave) representation. The dispersion relation
is then properly derivated and the implication of its
morphology are then discussed. Finally, in Section\il(\ref{Conc}),
brief concluding remarks follow.
\section{Two-dimensional axisymmetric dynamics}\label{Sec:II}
We now provide the basic equations governing the
two-dimensional axisymmetric dynamics of a plasma in
the Magneto-hydrodynamical representation. Having in mind
the specific application of such a dynamical system to
the morphology and stability problem of an accretion disk,
we write down the magnetic and velocity fields,
making use of the magnetic flux surface $\psi$ and
the angular velocity $\omega$, respectively, i.e.

\begin{eqnarray}
\vec{B} = - \frac{1}{r}\partial_z\psi \hat{e}_r
+\frac{\bar{B}_{\phi}}{r}\hat{e}_{\phi}
+ \frac{1}{r}\partial _r\psi \hat{e}_z
\label{a}\\
\vec{v} = \vec{v}_p + \omega r\hat{e}_{\phi}
\label{b}
\, .
\end{eqnarray}
Above, $\vec{v}_p = v_r\hat{e}_r + v_z\hat{e}_z$
is the poloidal component of the velocity field
and all the dynamical variables depend on $t$, $r$ and
$z$ only (the $\phi$ dependence being
suppressed because of the axial symmetry).

The dynamics of the two magnetic
variables $\psi$ and $\bar{B}_{\phi}$ can
be easily fixed by the ideal MHD equations
involving the magnetic structure of
the plasma. In particular, the azimuthal
component of the electron force balance
provides the equation

\begin{equation}
\partial _t\psi + \vec{v}_p
\cdot \vec{\nabla}\psi = 0
\,.
\label{psi}
\end{equation}

Analogously,  the azimuthal
component of the induction equation
yields the dynamics of $\bar{B}_{\phi}$ as

\begin{eqnarray}
\nonumber
\partial_t \bar{B}_{\phi} + \vec{v}_p \cdot
\vec{\nabla} \bar{B}_{\phi}  + \bar{B}_{\phi}\vec{\nabla} \cdot
\vec{v}_p=r(\partial_z\omega \partial_r \psi
- \partial_r\omega \partial_z\psi ).
\\\label{beb1}
\end{eqnarray}

The evolution of the angular velocity
$\omega$
and of the poloidal component
$\vec{v}_p$ is described by the
momentum conservation system,
whose azimuthal component reads

\begin{equation}
\rho r\left(\partial_t \omega + \vec{v}_p\cdot
\vec{\nabla} \omega \right)
+2\rho v_r\omega=\frac{1}{4\pi r^2}
\left(\partial_r\psi \partial_z\bar{B}_{\phi}
- \partial_z\psi \partial_r\bar{B}_{\phi} \right)
\, ,
\label{bei}
\end{equation}

while the poloidal ones provide

\begin{eqnarray}
\label{poeq}
\rho \left(\partial_t\vec{v}_p +
\vec{v}_p\cdot \vec{\nabla}\vec{v}_p
- \omega^2r\hat{e}_r\right)=- \vec{\nabla}p -
\nonumber \\
-\frac{1}{4\pi r}\left[ \partial_r\left(\frac{1}{r} \partial_r\psi\right)
+ \frac{1}{r}\partial^2_z\psi
\right] \vec{\nabla}\psi -
\nonumber \\
- \frac{1}{8\pi r^2}\vec{\nabla}\bar{B}_{\phi}^2 + \vec{F}^e_p
\,.
\end{eqnarray}
Above $\vec{F}^e_p$ is the external
poloidal force acting in the disk and,
in what follows, we will identify it with
the gravity of the central astrophysical
object.

The dynamics of the magnetized
plasma, as described in the ideal MHD is
then closed by providing the mass
conservation relation, i.e. the
continuity equation
for the mass density $\rho$

\begin{equation}
\partial_t\rho + \vec{\nabla}\cdot
\left(\rho \vec{v}_p\right)=0
\,
\label{coeq}
\end{equation}

and eventually assigning a
suitable equation of state to
express the pressure $p$, for
instance in the barotropic form
$p=p(\rho)$.
In the perturbation
analysis below, we will not need to
specify the plasma equation of state,
simply requiring its incompressibility.

We conclude by stressing how Eqs.\il
(\ref{beb1}) and (\ref{bei})
correlate the two azimuthal components
of the magnetic and velocity fields
respectively, making the gradient
of one as source term for the generation
of the other one (the left-hand-sides
of these equations are linear homogeneous
operators in the corresponding dynamical
variable).

\section{Background morphology}\label{Sec:mor}

We consider as background plasma
a purely rotating steady configuration
$\vec{v}_0 = \omega _0r
\hat{e}_{\phi}$ (here the
suffix $0$ denotes background quantities).
The rotation is clearly differential
across the disk, i.e.
$\omega _0 = \omega _0(r,z)$,
but the validity of the co-rotation
theorem \cite{F37} (holding for a
stationary axisymmetric plasma)
requires the condition
$\omega _0 = \omega _0(\psi _0)$,
$\psi _0$ being the background
magnetic surface. Accounting for
the validity of the co-rotation theorem
is the basic feature of our study of
the two-dimensional MRI and it constitute
the major difference with the analysis
in \cite{Balbus95}.

When referred to the background
equilibrium configuration, the dynamical
system discussed in the previous
section, reduces to a force balance
system-
The gravity of the central body,
around which the plasma disk is orbiting,
is crucial in fixing the steady
morphology, via the gravostatic
equilibrium equation,
involving also the background
mass density $\rho _0$ and pressure $p_0$.
Furthermore, since the plasma disk
is embedded in the vacuum magnetic field
of the central object, described via
the function $\psi _0$,  the
Lorentz force-free condition must hold.
Thus, we respectively get the two
equations

\begin{eqnarray}
\label{greq}
\vec{\nabla}p_0=\rho_0 \left(\omega_0^2(\psi_0)r\hat{e}_r -
\omega_K^2(r,z^2)\vec{r}_p\right) \\
\label{pore}
\frac{1}{4\pi r}\left[\partial_r\left(\frac{1}{r} \partial_r\psi_0\right)
+ \frac{1}{r}\partial^2_z\psi_0
\right]=0
\, ,
\end{eqnarray}

where $\omega _K$ denotes the
Keplerian frequency and $\vec{r}_p$
is the vector radius in the  meridian plane.

After solving the second of these equations
to get the form of $\psi _0$, for instance in
the form of a dipole contribution
(typical in compact astrophysical objects),
we can assign the function $\Omega _0(\psi )$
and the equation of state for the background plasma,
to determine the profile of the mass density
$\rho_0$ by the integration of the first one.
However, for the perturbation analysis,
here developed, such details are unessential (indeed we deal with a local
approach) and we
can directly proceed toward the linear dynamics.
Finally, we note that, while for
a thin configuration the disk angular
velocity is almost Keplerian
(i.e. $\omega _0 \simeq \omega _K$),
the radial pressure gradient takes
a relevant role for a thick disk
profile (i.e. $\omega _0\neq \omega _K$),
see \cite{1997MNRAS.288...63O}.

\section{Linear perturbation theory}\label{IV}

We now address the linear perturbation approach,
which is based on axisymmetric non-stationary
corrections to the equilibrium, also assumed of
very small spatial scale (their wavevectors have large
magnitude) with respect to the background quantities.
By other words, we deal with a local perturbation
approach, in which the linear terms, denoted
via the suffix $1$, are expanded in Fourier series,
analyzing the single monochromatic modes, i.e.

\begin{equation}
(...)_1(t, \vec{r}_p) =
\bar{(...)}_1\exp \left\{
i\left(\vec{k}\cdot\vec{r}_p - \Omega t\right) \right\}
\, \quad \, \bar{(...)}_1 = const
\, .
\label{la}
\end{equation}

Treating the perturbations as local corrections
($\mid \vec{k}\cdot\vec{r}_{0}\mid\ll 1$, being $r_0$ the fiducial
radius, around which the mode lives), does not means
that we deal with  a homogeneous background.
In fact, each background quantity is calculated
at the fiducial coordinates $\{ r_0, z_0\}$ and it
behaves as a constant coefficient in the perturbation scheme,
but this is true also for the spatial gradients
and, in particular, for the derivatives of the
background angular velocity
(which are expected to trigger the MRI).

Furthermore, we consider an incompressible plasma,
for which $\vec{\nabla}\cdot \vec{v} =
\vec{\nabla}\cdot \vec{v}_p = 0$ and a vanishing background azimuthal magnetic field. Often
(see \cite{BH98}), such a request comes out as the
so-called ``Boussinesq  approximation'', when a local
approach is pursued, but here it must be regarded
as a basic feature of the system, introduces to better
select the Alfvenic signature of the MRI.
This same point of view leads us
to simplify our analysis, by
eliminating the magnetic pressure
with the request that the wavevector
$\vec{k}$ be parallel to the background
magnetic field $\vec{B}_0$, i.e.
we require
$\vec{k}\cdot \vec{\nabla}\psi _0 = 0$.

Since each monochromatic mode obeys the
relations
\begin{equation}
\partial _t(...)_1 = -i\Omega (...)_1
\, ; \,
\vec{\nabla}(...)_1 = i\vec{k}(...)_1
\, ,
\label{dlt}
\end{equation}

we can easily restate the dynamics of the
perturbations (as deduced by the basic system of the ideal
MHD evolution equations) in terms of a closed
algebraic system, providing the dispersion relation
for the mode spectrum.

We observe how the poloidal velocity $\vec{v}_p$
is absent in the background equilibrium and it is
therefore natural to express it by means of the
poloidal shift vector $\vec{\xi}_p$ of the
plasma elements, i.e. we have

\begin{equation}
\vec{v}_p = \partial _t\vec{\xi}_p =
-i\Omega \vec{\xi}_p
\Rightarrow \vec{\nabla}\cdot \vec{\xi}_p= i\vec{k}\cdot \vec{\xi}_p = 0
\, .
\label{in}
\end{equation}

First of all, we stress how the continuity
equation (\ref{coeq}) is reduced by the incompressibility
constraint above to the vanishing nature of the
perturbed mass density $\rho _1$. Furthermore,
Eq. (\ref{psi}) provides the perturbed magnetic
surface function $\psi _1$ in the form

\begin{equation}
\psi _1 = - \vec{\xi}_p \cdot \vec{\nabla}\psi _0
\, .
\label{beae}
\end{equation}

Hence, the poloidal component (\ref{poeq}) of the momentum conservation system
reads, to the linear approximation, as follows

\begin{equation}
\Omega ^2 \vec{\xi}_p +
2\omega _0\left( \dot{\omega}_0\psi _1 +
\omega ^+_1\right) \hat{e}_r
= i\vec{k}\frac{p_1}{\rho _0} - \frac{k^2\psi _1}{4\pi
r^2\rho _0} \vec{\nabla}\psi _0
\, .
\label{beqk}
\end{equation}

Above, we split the perturbed angular velocity
in terms of its first order co-rotation
contribution $\dot{\omega}_0\psi _1$
(where $\dot{\omega}_0 \equiv (d\omega /d\psi)_{\psi = \psi _0}$)
and a generic linear deformation $\omega _1^+$.

Taking the scalar product of the system above with the
wavevector $\vec{k}$, we easily get the perturbed
pressure contribution as a consequence of preserving the
incompressibility condition during the dynamics, i.e.

\begin{equation}
k^2p_1
	=-
2i\rho_0\omega _0k_rr\left(
\dot{\omega}_0 \psi _1 + \omega ^+_1\right)
\, .
\label{rho1}
\end{equation}

The scalar product of Eq. (\ref{beqk}) with
$\vec{\nabla}\psi _0$, once Eq. (\ref{beae}) is
used, yields the basic relation

\begin{equation}\label{qfef}
\left(\Omega^2 - y_r - \omega_A^2\right)\psi _1
=2\omega_0r\partial_r\psi_0 \omega^+_1
\, ,
\end{equation}

where $\omega _A^2 \equiv
(\vec{k}\cdot \vec{B}_0)^2/ 4\pi
\rho _0 = k^2v_A^2$ ($v_A$ being
the Alfven velocity) and
$y_r\equiv 2\omega _0r\partial _r\omega _0$.

For a monochromatic mode the linearized
Eqs. (\ref{beb1}) and (\ref{bei})
take respectively the form

\begin{eqnarray}
\label{psx1}
r\Omega \omega^+_1=-2\omega_0\Omega \xi_r
- \frac{\vec{k}\cdot \vec{B}_0}{4\pi \rho_0r}(\bar{B}_{\phi})_1 \end{eqnarray}
and
\begin{eqnarray}
\label{eeb1}
\Omega (\bar{B}_{\phi})_1=- r^2\vec{k}\cdot \vec{B}_0\omega^+_1
\, ,
\end{eqnarray}

which combined together give
the relation

\begin{equation}
r\left(\Omega^2 - \bar{\omega}_A^2\right)\omega^+_1
=- 2\omega_0\Omega^2\xi_r
\, .
\label{omxi}
\end{equation}

Finally, the radial component
of Eq. (\ref{beqk}), using Eq.
(\ref{rho1}), can be restated as

\begin{equation}
\Omega^2\partial_r\psi_0\xi_r=-\left(\alpha y_r + v_{Az}^2k^2\right) \psi_1
-2\alpha \omega_0r\partial_r\psi_0 \omega^+_1
\, ,
\label{xiom}
\end{equation}

where $\alpha \equiv 1 - k_r^2/k^2$.

The system of Eqs. (\ref{qfef}),
(\ref{omxi}) and (\ref{xiom}) is
a closed algebraic set in the variables
$\psi _1$, $\omega ^+_1$ and $\xi _r$
and they can be easily combined to
obtain the following dispersion
relation

\begin{eqnarray}
\nonumber
&&\Omega ^4-{ b}\;\Omega^2
+ { c}  = 0,
\\
&& \label{dire}{ c}\equiv \omega _A^2\left( y_r + \omega _A^2\right)\quad{{b}}\equiv \left( K_0^2 + 2\omega _A^2
\right) + 4\omega _0^2 (\alpha-1),
\end{eqnarray}

Here

\begin{equation}
K^2_0\equiv
\frac{1}{r^3}\partial _r
\left(r^4\omega _0^2\right) =
y_r + 4\omega _0^2.
\label{k0}
\end{equation}

It is immediate to check that the
necessary condition to get MRI is
provided by the inequality
$\omega _A^2 < - y_r$, which ensures
 $c < 0$. Indeed, also
the request $b < 0$ could provide
unstable behaviors, but the expression
of $K_0^2$ implies that
$b = c / \omega _A^2
+ 4\omega _0^2(k_z^2/k^2)$ and
hence $b < 0$ requires again $c < 0$.

Thus, we see how, when we take into
account the validity of the
co-rotation theorem, the condition
for MRI is the same in the case of a
stratified thick disk as in a thin
disk configuration \cite{BH91}.
Such an issue relies on the cancellation
of the $z$-derivative of the angular
velocity in the dispersion relation
and this fact is the main difference
between the present analysis and
that one performed in \cite{Balbus95},
where a vector formulation is addressed
instead using the magnetic flux surface
function (regardless of the co-rotation
theorem).

\section{Concluding remarks}\label{Conc}
We analyzed the stability of a stratified and
differently rotating ideal plasma disk,
as described in a two-dimensional axisymmetric scheme.
After the set up of the basic dynamical system, required
to address the considered problem, we briefly characterize
the morphology of a the steady background configuration
of the plasma disk.  The steady  configuration is assumed  to be embedded in the gravitational and the magnetic
field of the central object, according to the
astrophysical implementation of our analysis in
the behavior and stability of stellar accretion structures.
Then, we constructed the Fourier representation of the
linear perturbation dynamics, allowed by the request to
deal with short wave-length deformation of the background
and hence by a local perturbation approach.
The combination of such algebraic equations lead to
the morphology of the dispersion relation, providing
the structure of the spectral modes.
As result of such a procedure, we arrive to study the
profile of the MRI in the case of a stratified disk,
evaluated with the help of two simplifying assumptions
for selecting the Alfvenic nature of this instability:
the plasma is required to be incompressible
(so removing the acoustic modes) and the perturbations
propagate along the background magnetic field
(i.e. we deal with zero first order magnetic pressure).
As basic issue of our analysis we demonstrate how,
accounting for the co-rotation theorem on the background,
has significant implications on the linear dispersion relation.
In particular, the co-rotation theorem is at the ground
of the cancelation of the vertical derivative of
the background disk angular velocity from the structure
of the linear unstable modes. Such a feature does not
emerge in the vector approach depicted in
\cite{Balbus95}, where the co-rotation profile is not
taken into account.
Thus, the dispersion relation for a stratified thick
accretion disk retains the same structure as in the
thin disk approximation, i.e. the condition to trigger
the MRI is associated to the same inequality.
As a result the stability of the disk, the
corresponding turbulence and the associated angular
momentum transport are essentially determined by the
radial profile of the background configuration,
which is fixed by the balance between the gravitational
force, the pressure gradient and the centripetal force.
The main difference in the MRI morphology between thin
and thick accretion disks, consists in the role that the
radial pressure can take in the later case, while
in the former the disk angular frequency has
essentially a Keplerian profile.

\bibliographystyle{jpp}

\bibliography{jpp-instructions}

\end{document}